\documentclass[twocolumn, prl,showpacs,superscriptaddress]{revtex4}

\usepackage{amsmath}
\usepackage{mathrsfs} 
\usepackage{amssymb} 
\usepackage{graphicx}
\usepackage{subfigure}
\usepackage{dcolumn}
\usepackage{bm}
\usepackage{natbib}

\begin{document}

\title{Electron-Phonon Coupling in Two-Dimensional Silicene and Germanene}
\date{\today}
\author{Jia-An Yan}
\email{jiaanyan@gmail.com}
\affiliation{Department of Physics, Astronomy, and Geosciences, Towson University, 8000 York Road, Towson, MD 21252, USA}
\author{Ryan Stein}
\affiliation{Department of Physics, Astronomy, and Geosciences, Towson University, 8000 York Road, Towson, MD 21252, USA}
\author{David M. Schaefer}
\affiliation{Department of Physics, Astronomy, and Geosciences, Towson University, 8000 York Road, Towson, MD 21252, USA}
\author{Xiao-Qian Wang}
\affiliation{Department of Physics and Center for Functional Nanoscale Materials, Clark Atlanta University, Atlanta, Georgia 30314, USA}
\author{M. Y. Chou}
\email{meiyin.chou@physics.gatech.edu}
\affiliation{School of Physics, Georgia Institute of Technology,
Atlanta, Georgia 30332, USA}
\affiliation{Institute of Atomic and Molecular Sciences, Academia Sinica, Taipei 10617, Taiwan}

\begin{abstract}
Following the work in graphene, we report a first-principles study of electron-phonon coupling (EPC) in low-buckled (LB) monolayer silicene and germanene. Despite of the similar honeycomb atomic arrangement and linear band dispersion, the EPC matrix-element squares of the $\Gamma$-$E_g$ and K-$A_1$ modes in silicene are only about 50\% of those in graphene. However, the smaller Fermi velocity in silicene compensates this reduction by providing a larger joint electronic density of states near the Dirac point. We predict that Kohn anomalies associated with these two optical modes are significant in silicene. In addition, the EPC-induced frequency shift and linewidth of the Raman-active $\Gamma$-$E_g$ mode in silicene are calculated as a function of doping. The results are comparable to those in graphene, indicating a similar non-adiabatic dynamical origin. In contrast, the EPC in germanene is found to be much reduced.\\\
\end{abstract}

\maketitle

The success in fabricating graphene in the laboratory has suggested many opportunities in producing other two-dimensional (2D) materials \cite{Neto2009,Rogers2011,Radisavljevic2011}. Among them, the possibility of creating a monolayer honeycomb lattice made of Si or Ge (namely, silicene or germanene) is of particular interest because of the compatibility with current silicon technology. With the same $s^2p^2$ electronic configuration in group-IV elements, silicene and germanene have been predicted to exhibit similar linear band dispersions near the Fermi level \cite{Takeda1994,Cahangirov2009}. Charge carriers at low energies therein also behave like massless Dirac Fermions. In contrast to the planar structure of graphene, the stable structure for silicene and germanene is low-buckled (LB) due to their larger atomic radius \cite{Takeda1994,Cahangirov2009,Verri2007}, leading to an electrically tunable band gap \cite{Ni2012,Drummond2012} and topological phase transition \cite{Ezawa2012}. Recent progresses \cite{Fleurence2012,Vogt2012,Chen2012,Feng2012,Jamgotchian2012,Lin2012} in the growth of silicene and germanene have paved the way for further studying practical device applications and low-dimensional physics in these systems.

Electron-phonon coupling (EPC) is critical for understanding many phenomena in carbon-based materials,
including high field transport \cite{Yao2000}, phonon renormalization \cite{Das2009}, Raman spectra \cite{Ferrari2006, Yan2007}, and Kohn anomalies \cite{Piscanec2004,Si2012}. The Kohn anomaly \cite{Kohn1959} is an anomalous feature of phonon dispersions in metals and semimetals arising from a strong EPC for specific phonon modes. In graphite and 2D graphene, Kohn anomalies associated with the highest-branch optical $\Gamma$-$E_{2g}$ and K-$A_1'$ modes have been reported \cite{Piscanec2004}. In addition, it has also been demonstrated that the frequency of the $\Gamma$-$E_{2g}$ mode in graphene varies non-adiabatically as a function of doping \cite{Lazzeri2006}. Given the increased attention to silicene and germanene, it would be of fundamental interest to see if these anomalous features in graphene carry over to other 2D systems with a similar honeycomb lattice structure and linear band dispersions at low energies. It is expected that the LB atomic arrangement in silicene and germanene would more or less affect the bonding characteristics between nearest-neighbor (NN) atoms. Consequently, the delicate interplay between electrons and phonons may be significantly modified. Understanding EPC in silicene and germanene from first principles is highly desirable and essential in order to facilitate their further characterization and applications.

\begin{figure}[tbp]
\centering
   \includegraphics[width=9cm,clip]{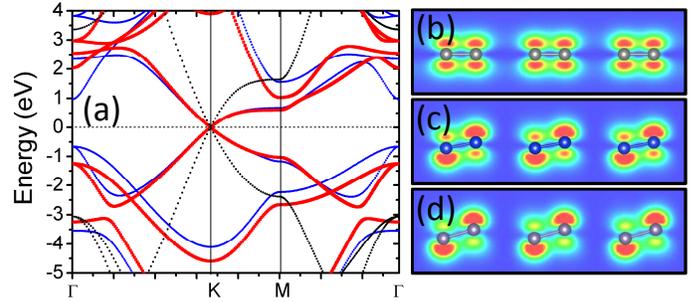}
 \caption{(Color online) Band dispersions of silicene (red solid) and germanene (blue solid). The band structure of graphene (dotted black) are also shown for comparison. (b)-(d) The band-decomposed charge density distribution of the $\pi$ ($\pi^*$) bands near the Fermi level on the \{110\} plane in graphene, silicene, and germanene, respectively \cite{Vesta2011}.  The Fermi level is shifted to zero in (a). }\label{fig:band}
\end{figure}

In this Letter, we show that there exist two kinks in the phonon dispersions of the LB silicene and germanene, corresponding to the highest-branch optical $\Gamma$-$E_g$ and $K$-$A_1$ modes, respectively, implying Kohn anomalies in silicene and germanene, although the anomaly is much weaker in the latter. A further frozen-phonon calculation found that the EPC matrix-element squares of these two modes in silicene are merely 50\% of those in graphene. However, the phonon linewidth of the $\Gamma$-$E_g$ mode due to the EPC is comparable with that in graphene. This can be explained by the compensating effect of the smaller linear-band slope (i.e., the Fermi velocity $v_F$) in silicene. In contrast, the EPC matrix elements in germanene are 10 times smaller than those in silicene, indicating a much reduced EPC in this particular 2D system.

\begin{figure*}[tbp]
\centering
 \includegraphics[width=15cm,clip]{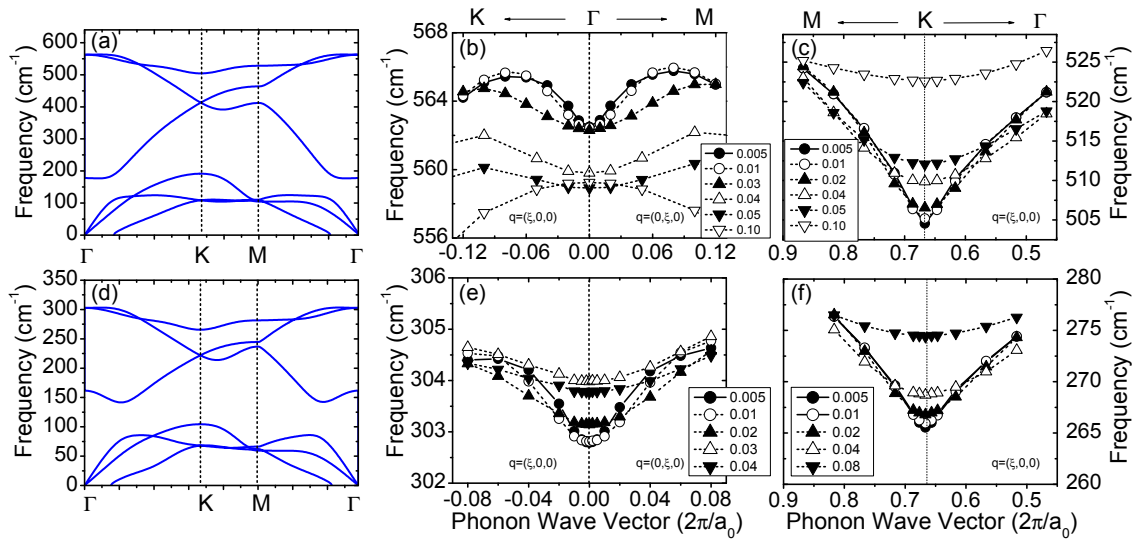}
 \caption{(Color online) Phonon dispersions of (a) silicene, with the highest optical modes around (b) $q$ = $\Gamma$ and (c) $q$ = K as a function of smearing parameter $\sigma$. Phonon dispersions of (d) germanene, with the highest optical modes around (e) $q$ = $\Gamma$ and (f) $q$ = K as a function of smearing parameter $\sigma$. The calculated frequencies are marked, with the lines being guides to the eye.}\label{fig:kohn}
\end{figure*}

Our calculations are performed within density functional theory (DFT) with the local density approximation (LDA). For the phonon calculations we employ density functional perturbation theory (DFPT) \cite{Baroni2001} as implemented in the Quantum ESPRESSO code \cite{pwscf}. Norm-conserving pseudopotentials \cite{Troullier1991} for Si and Ge are adopted to describe the core-valence interaction. The wave functions of the valence electrons are expanded in plane waves with a kinetic-energy cutoff of 36 Ry and 40 Ry for silicene and germanene, respectively. A vacuum region of 20 \AA~ is introduced to eliminate any artificial interaction between neighboring supercells along the perpendicular direction. The relaxed lattice constants are $a_0$ = 3.83 \AA~ for silicene and 3.95 \AA~ for germanene. Accordingly, the NN bond lengths are equal to 2.25 \AA~and 2.37 \AA, nearly 4.3\% and 3.3\% smaller than the corresponding values in bulk Si and Ge, respectively. The relaxed buckled separations are $\Delta$ = 0.46 \AA~for silicene and 0.64 \AA~for germanene. These results are in good agreement with previous DFT calculations \cite{Cahangirov2009}.

In Fig. 1(a) we show the calculated band structures of silicene (red) and germanene (blue). The band dispersions of graphene (black) are also shown for comparison. Clearly, the linear bands cross at the Fermi level $E_F$ at $q$ = $K$, indicating a semi-metallic nature for both LB silicene and germanene. By fitting the linear $\pi$ and $\pi^*$ bands at $k$ = $K$ + $k'$ to the expression: $E(k') = \hbar v_F k' = \beta k'$, we obtained the Fermi velocity $v_F$ = 0.54$\times$10$^6$ m/s for silicene and 0.53$\times$10$^6$ m/s for germanene, respectively. These values are only 63\% of the LDA value of 0.86$\times$10$^6$ m/s in graphene, consistent with previous plane-wave DFT calculations \cite{Bechstedt2012}.

One important feature of silicene and germanene is their LB structure. To see the effects of buckling on the bonding characteristics, Figs.~1(b)-(d) show the charge density distribution for the linear bands from graphene to germanene. In graphene, the carbon $s$ orbital hybridizes with $p_x$ and $p_y$ atomic orbitals to form $sp^2$ hybrids. This hybridization leads to a strong $\sigma$ bonding between NN carbon atoms and keeps them in a plane. The perpendicular $p_z$ orbitals gives rise to $\pi$ and $\pi^*$ bands, as shown in Fig.~1(b). As the bond length between two NN atoms increases, the $sp^2$ orbitals become weaker and the $\sigma$ bonds are not strong enough to keep Si or Ge atoms in the same plane, leading to a low-buckled structure. One sublattice moves upward by $\Delta/2$, while the other moves downward by $\Delta/2$, with $\Delta$ = 0.46 and 0.64 \AA~ for silicene and germanene, respectively. This buckling brings part of the $p_z$ orbital closer to neighboring $sp^2$ orbitals, creating further hybridization in silicene and germanene, as shown in Figs.~1(c) and (d), respectively.  As will be discussed below, these bonding characteristics significantly affect EPC in silicene and germanene compared to graphene.


\begin{table*}[tbp]
 \caption{Calculated frequency $\omega$ (in cm$^{-1}$), linewidth $\gamma$ (in cm$^{-1}$), $\langle g_q^2 \rangle_\texttt{F}$ (in eV$^2$) for the highest optical modes at $\Gamma$ and $K$ in silicene and germanene. Corresponding results of graphene are also listed for comparison.} \label{tab:freq}
\begin{tabular}{llrrrrrrr}
\hline
            &  & \multicolumn{3}{c}{$q$ at $\Gamma$}  & \multicolumn{3}{c}{$q$ at K} & \\
            \cline{3-5} \cline{6-8}
            &  & $\omega$ & $\gamma$ &$\langle g_\Gamma^2 \rangle_\texttt{F}$ & $\omega$ & $\gamma$ & $\langle g_K^2 \rangle_\texttt{F}$  & $\frac{\langle g_K^2 \rangle_\texttt{F} \omega_K}{\langle g_\Gamma^2 \rangle_\texttt{F} \omega_\Gamma}$\\
            \hline
Graphene   & Ref. [\citenum{Piscanec2004}] & 1540 & 11.2 & 0.0405 & 1250 &21.8$^{a}$ & 0.0994 & 1.98 \\
           & Ref. [\citenum{me2009}] & 1586 & 11.2 & 0.0401 & 1320 & 20.4 & 0.0986 & 2.05 \\
           \hline
Silicene   & This work & 562 & 13.3 & 0.0223  & 506 & 21.5 & 0.0478  & 1.92 \\
           \hline
Germanene  & This work & 303 & 0.6 & 0.00214  & 267 & 1.2 & 0.00464  & 1.91 \\
\hline
$^{a}$Ref. \citenum{Park2008}.
\end{tabular}
\end{table*}

Now we turn to the phonon properties in these systems. The calculated phonon dispersions of silicene and germanene are shown in Figs.~2(a) and (d), respectively. Due to the buckling, the point group symmetry of silicene and germanene is reduced to $D_{3d}$. The highest optical mode at $\Gamma$ belongs to a doubly degenerate $E_g$ mode with a frequency of $\omega_\Gamma$ = 562 cm$^{-1}$ for silicene and 303 cm$^{-1}$ for germanene. The $E_g$ mode in silicene is Raman active, nearly 50 cm$^{-1}$ higher than the highest optical mode (510 cm$^{-1}$ in LDA) in bulk silicon. At the BZ corner $q$ = $K$, the highest optical mode belongs to the representation of $A_1$, with a frequency of $\omega_K$ = 506 cm$^{-1}$ for silicene and 267 cm$^{-1}$ for germanene. These results are listed in Table 1.

The nonanalytic behavior of the phonon dispersion at $\Gamma$ and $K$ can be examined by varying the fictitious electronic smearing temperature $\sigma$ in the direct DFPT phonon calculations \cite{Piscanec2004}. The smearing temperature $\sigma$ alters the occupation of the electronic states near the Fermi level, and the discontinuities can be obtained by taking the limit $\sigma$ $\rightarrow$ 0. Figs.~2(b)-(c) and 2(e)-(f) show the evolution of the dispersions near $\Gamma$-$E_g$ and $K$-$A_1$ modes for silicene and germanene, respectively. The results for $\sigma$ = 0.01 Ry and $\sigma$ = 0.005 Ry are similar, providing a proper $\sigma \rightarrow$ 0 limit on the scale of the figure. As shown in Fig.~2, these discontinuities in the frequency derivative are smoothed out and disappear quickly with increasing $\sigma$. It is thus confirmed that these discontinuities result from an anomalous screening of the electrons around the Fermi energy, with the visible kinks in these dispersions at $\Gamma$ and K corresponding to Kohn anomalies for both the $E_g$ and $A_1$ modes.

Comparing the phonon dispersions in Fig.~2 with those for graphene \cite{Piscanec2004}, we see that the range of frequency variation ($\sim$ 6 cm$^{-1}$ for silicene and $\sim$ 2 cm$^{-1}$ for germanene at $\Gamma$) when the smearing parameter is changed is significantly smaller than that in graphene ($\sim$ 40 cm$^{-1}$), implying a much weaker EPC in silicene and germanene. To make a quantitative comparison, we have calculated the EPC matrix elements for these optical modes. The EPC matrix element is given by
\begin{eqnarray}
g_{(\mathbf{k+q})m,\mathbf{k}n}^{\nu}=\sqrt{\frac{\hbar}{2M\omega_{\mathbf{q}}^{\nu}}}\langle
m\;\mathbf{k+q}|\frac{\delta V_{scf}}{\delta
u_{\mathbf{q}}^{\nu}}|n\;\mathbf{k}\rangle,
\end{eqnarray}
where $\delta V_{scf}\equiv
V_{scf}(u_{\mathbf{q}}^{\nu})-V_{scf}(0)$ is the variation of the self-consistent potential due to the perturbation of a phonon displacement with wave vector $\mathbf{q}$, frequency $\omega$, and branch index $\nu$. $|n\;\mathbf{k}\rangle$ is the $n$th electronic state at wave vector $\mathbf{k}$. Details of the calculation can be found in Ref.~\citenum{me2009}.

We first compare the square of the EPC matrix elements in silicene with those in graphene. Since at $k$ = K, the electronic states are doubly degenerate, we calculate the average EPC matrix-element square over the Fermi surface defined as $\langle g_q^2\rangle_F = \sum_{i,j}^\pi |g_{(K+q)i,Kj}|^2/4$ with $q$ = $\Gamma$ or K, where the sum is performed over the two degenerate $\pi$ bands at $E_F$. The results are summarized in Table~1. As shown by Piscanec et al. \cite{Piscanec2004}, the strength of the Kohn anomalies in graphene is linearly proportional to this average. In silicene, the calculated values are $\langle g_K^2\rangle_F = \sum_{i,j}^\pi |g_{(2K)i,Kj}|^2/4$ = 0.0478 eV$^2$ and $\langle g_\Gamma^2\rangle_F = \sum_{i,j}^\pi |g_{Ki,Kj}|^2/4$ = 0.0223 eV$^2$. These values are roughly about 50\% smaller than those in graphene (0.0986 eV$^2$ and 0.0401 eV$^2$, respectively \cite{me2009}). The smaller EPC matrix-element squares indicate weaker Kohn anomalies, consistent with the direct phonon calculations as discussed above. Surprisingly, the calculated EPC matrix-element square in germanene is nearly 10 times smaller than that in silicene, implying a negligible EPC in germanene. These differences in the EPC matrix elements reflect the distinct bonding characteristic in graphene, silicene and germanene. An interesting finding is that in both silicene and germanene the ratio
\begin{equation}
\frac{\langle g_K^2 \rangle_F \omega_K} {\langle g_\Gamma^2 \rangle_F \omega_\Gamma} = 1.9 \approx 2
\end{equation}
is slightly smaller than the value of 2.05 in graphene. (A tight-binding model with NN interaction yields an exact ratio of 2 \cite{Piscanec2004}.)

In the following, we focus on the electron-phonon interaction as a function of doping in silicene. Since the $\Gamma$-$E_g$ mode in silicene is Raman active, the phonon frequency shift and the broadening as a function of the doping level can be measured by Raman spectroscopy. We start from the self-energy $\Pi_{\mathbf{q}\nu} (\omega)$ of a phonon with wave vector $\mathbf{q}$, branch index $\nu$, and frequency $\omega_{\mathbf{q}\nu}$. This quantity provides the renormalization and damping of the specific phonon due to the interaction with other elementary excitations. This method has been applied to study the optical phonon anomaly in bilayer graphene at ultrahigh doping levels \cite{me2012}. Within the Migdal approximation, the self-energy induced by the EPC can be expressed as \cite{Ando2007}:
\begin{equation}
\Pi_{\mathbf{q}\nu}(\omega)
=
2 \sum_{mn}\int\frac{d\mathbf{k}}{\Omega_{\mathrm{BZ}}}|g_{(\mathbf{k+q})m,\mathbf{k}n}^{\nu}|^2 \frac{[f(\epsilon_{m\mathbf{k+q}})-f(\epsilon_{n\mathbf{k}})]}{\epsilon_{m\mathbf{k+q}}-\epsilon_{n\mathbf{k}}-\hbar\omega-i\eta},
\end{equation}
where $\epsilon_{n\mathbf{k}}$ is the energy of an electronic state $|n\mathbf{k}\rangle$ with crystal momentum $\mathbf{k}$ and band index $n$, $f(\epsilon_{n\mathbf{k}})$ the corresponding Fermi occupation, and $\eta$ a positive infinitesimal. For a given mode $\omega=\omega_0$, the phonon linewidth is $\gamma$ = $-2\,\mathrm{Im}(\Pi_{\mathbf{q}\nu} (\omega_0))$ and the phonon frequency shift is $\Delta \omega$ = $[\mathrm{Re}(\Pi_{\mathbf{q}\nu} (\omega_0)|_{E_F}-\Pi_{\mathbf{q}\nu}(\omega_0)|_{E_F=0})]/\hbar$. We have carried out DFT calculations on a dense 201$\times$201 $k$-grid within a minizone (0.2$\times$0.2) enclosing the BZ corner $K$ in the reciprocal space. This is equivalent to 1000$\times$1000 $k$-grid sampling in the whole Brillouin zone. By changing the Fermi level $E_F$, we can investigate the dependence of $\gamma$ and $\Delta \omega$ on the doping level, assuming that the EPC matrix elements are unchanged. For all the linewidths calculated below, we used a parameter $\eta$ = 5 meV.

\begin{figure}[tbp]
\centering
  \includegraphics[width=8.5cm]{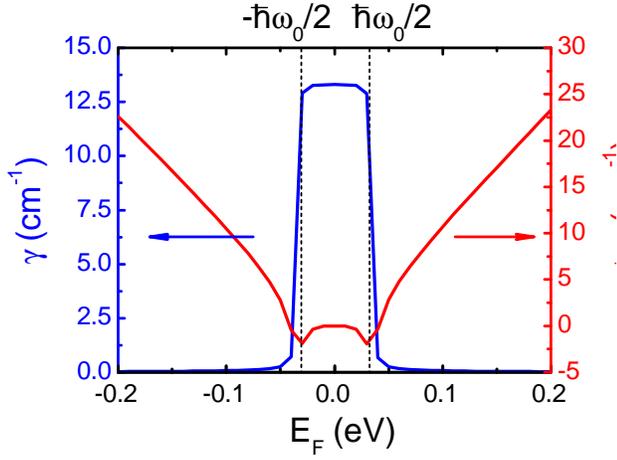}
 \caption{(Color online) Calculated phonon linewidth $\gamma$ (left axis) and frequency shift $\Delta \omega$ (right axis) in silicene as a function of the Fermi level $E_F$ for the $\Gamma$-$E_g$ mode with $\hbar \omega_0$ = 70 meV. The neutrality point has been shifted to zero. }\label{fig:gamma}
\end{figure}

Fig.~\ref{fig:gamma} shows the calculated linewidth $\gamma$ (left axis) of the $\Gamma$-$E_g$ mode as a function of the Fermi level $E_F$. In the low doping regime with $|E_F| < \hbar \omega_0/2 \sim 35$ meV, the $E_g$ mode can have a resonant coupling with the electron-hole pair from the the valence and conduction bands. As a result, $\gamma$ $\sim$ 13.5 cm$^{-1}$, a constant within this doping range. Surprisingly, the linewidth of silicene is slightly larger than that in graphene in spite of the relatively smaller EPC matrix element. This is mainly due to the different slopes of the linear bands between graphene and silicene. Fig.~4 schematically illustrates this effect. When the band is steeper, fewer states in the $k$ space satisfying the energy conservation will be allowed to couple with the phonon mode. In contrast, when the band is flatter, more electronic states will satisfy the energy conservation condition, giving rise to a large contribution to the phonon linewidth.

\begin{figure}[tbp]
\centering
  \includegraphics[width=8.5cm,clip]{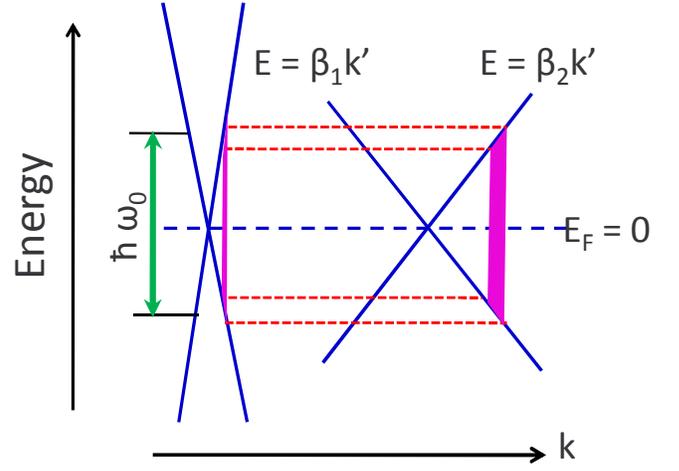}
 \caption{(Color online) Schematic plot of the effect of the band slope on the phonon linewidth. With $\beta_1 >$ $\beta_2$, system 1 has fewer states in the $k$ space that satisfy the energy conservation and are allowed to couple with the phonon mode than system 2, resulting in a smaller contribution to the self energy calculation in Eq.~(3).}\label{fig:scheme}
\end{figure}

In bulk silicon, the first order Raman peak is at 520 cm$^{-1}$, with a linewidth of $\gamma$ = 4.6 cm$^{-1}$ \cite{Parker1967}. This is mainly due to the anharmonicity effect \cite{Debernardi1995}, implying weak EPC. In contrast, the Raman active mode in silicene is at 562 cm$^{-1}$, nearly 40 cm$^{-1}$ higher than the experimental value in bulk. This is understandable, since in silicene, due to the reduced dimensionality, the Si-Si bond length of 2.25 \AA~ is about 4.2\% smaller than that in bulk silicon. A significant change on the phonon frequency is therefore expected. Moreover, the linewidth of the $E_g$ mode in silicene is about 13.5 cm$^{-1}$, implying a much enhanced EPC in silicene due to the reduced dimensionality.

The corresponding frequency shift $\Delta \omega$ is also presented in Fig.~\ref{fig:gamma} (right axis). With the increase of $|E_F|$, the $E_g$ mode slightly softens and reaches the singularity at $|E_F|$ = $\hbar \omega_0$/2 for both electron and hole dopings. Further increase of $|E_F|$ results in a hardening of the mode frequency. In particular, $\Delta \omega$ increases almost linearly as $|E_F|$ $\gg$ $\hbar \omega_0$/2. These results indicate that similar Raman features observed in graphene \cite{Yan2007} may also be observable in silicene.

In summary, we have shown that low-buckled silicene and germanene exhibit distinct electron-phonon coupling compared to graphene. Although the EPC matrix elements in silicene are much smaller than those in graphene, the smaller slope of the linear bands compensates this effect, leading to a comparable phonon linewidth of the $\Gamma$-$E_g$ mode as in graphene. Our finding implies that Raman spectroscopy may also serve as a power tool in silicene research. Furthermore, we have identified possible Kohn anomalies associated with $\Gamma$-$E_g$ and $K$-$A_1$ modes in silicene. In contrast, the EPC in germanene is found to be weak and nearly negligible.


J.A.Y. and R.S. acknowledge the support from SET at Towson University. J.A.Y. and X.Q.W. acknowledge
NSF PREM DMR-0934142 for support. M.Y.C. acknowledges support by the US Department of Energy, Office of Basic Energy Sciences, Division of Materials Sciences and Engineering under Award No. DEFG02-97ER45632. This research used computational resources at the National Energy Research Scientific Computing Center (supported by the Office of Science of the U.S. Department of Energy under Contract No. DE-AC02-05CH11231).

%
%


\begin{thebibliography}{99}

\bibitem{Neto2009} A. H. Castro Neto, F. Guinea, N. M. R. Peres, K. S. Novoselov and A. K. Geim, Rev. Mod. Phys. \textbf{81}, 109-162 (2009) and references therein.

\bibitem{Rogers2011} J. A. Rogers, M. G. Lagally, and R. G. Nuzzo, Nature \textbf{477}, 45-53 (2011).

\bibitem{Radisavljevic2011} B. Radisavljevic, A. Radenovic, J. Brivio, V. Giacometti and A. Kis, Nat. Nanotech. \textbf{6}, 147-150 (2011).

\bibitem{Takeda1994} K. Takeda and K. Shiraishi, Phys. Rev. B \textbf{50}, 14916-14922 (1994).

\bibitem{Cahangirov2009} S. Cahangirov, M. Topsakal, E. Akt\"{u}rk, H. Sahin, and S. Ciraci, Phys. Rev. Lett. \textbf{102}, 236804 (2009).

\bibitem{Verri2007} G. G. Guzm\'{a}n-Verri, L. C. Lew Yan Voon, Phys. Rev. B \textbf{76}, 075131 (2007).

\bibitem{Ni2012} Z. Ni, Q. Liu, K. Tang, J. Zheng, J. Zhou, R. Qin, Z. Gao,
D. Yu, and J. Lu, Nano Lett. \textbf{12}, 113-118 (2012).

\bibitem{Drummond2012} N. D. Drummond, V. Z\'{o}lyomi, and V. I. Fal'ko, Phys. Rev. B \textbf{85}, 075423 (2012).

\bibitem{Ezawa2012} M. Ezawa, Phys. Rev. Lett. \textbf{109}, 055502 (2012).


\bibitem{Vogt2012} P. Vogt, P. DePadova, C. Quaresima, J. Avila, E. Frantzeskakis, M. C. Asensio, A. Resta, B. Ealet,
and G. LeLay, Phys. Rev. Lett. \textbf{108}, 155501 (2012).

\bibitem{Fleurence2012} A. Fleurence, R. Friedlein, T. Ozaki, H. Kawai, Y. Wang, and Y. Yamada-Takamura, Phys. Rev. Lett. \textbf{108}, 245501 (2012).

\bibitem{Chen2012} L. Chen, C.-C. Liu, B. Feng, X. He, P. Cheng, Z. Ding, S. Meng, Y. Yao, K. Wu, Phys. Rev. Lett. \textbf{109}, 056804 (2012).

\bibitem{Feng2012} B. Feng, Z. Ding, S. Meng, Y. Yao, X. He, P. Cheng, L. Chen, and K. Wu, Nano Lett., \textbf{12}, 3507-3511 (2012).

\bibitem{Jamgotchian2012} H. Jamgotchian, Y. Colignon, N. Hamzaoui, B. Ealet, J. Y. Hoarau, B. Aufray, and J. P. Bib\'{e}rian, J. Phys. Condens. Matter \textbf{24}, 172001 (2012).


\bibitem{Lin2012} C.-L. Lin, R. Arafune, K. Kawahara, N. Tsukahara, E. Minamitani, Y. Kim, N. Takagi, and M. Kawai, App. Phys. Exp. \textbf{5}, 045802 (2012).

\bibitem{Yao2000} Z. Yao, C. L. Kane, and C. Dekker, Phys. Rev. Lett. \textbf{84}, 2941-2944 (2000).
%
\bibitem{Das2009} A. Das, B. Chakraborty, S. Piscanec, S. Pisana, A. K. Sood, and A. C. Ferrari, Phys. Rev. B \textbf{79}, 155417 (2009).

\bibitem{Ferrari2006} A. C. Ferrari, J. C. Meyer, V. Scardaci, C. Casiraghi, M. Lazzeri, F. Mauri, S. Piscanec, D. Jiang, K. S. Novoselov, S. Roth, and A. K. Geim, Phys. Rev. Lett. \textbf{97}, 187401 (2006).

\bibitem{Yan2007} J. Yan, Y. Zhang, P. Kim, A. Pinczuk, Phys. Rev. Lett. \textbf{98}, 166802 (2007).


\bibitem{Piscanec2004} S. Piscanec, M. Lazzeri, F. Mauri, A. C. Ferrari, and J. Robertson, Phys. Rev. Lett. \textbf{93}, 185503 (2004).

\bibitem{Si2012} C. Si, W. Duan, Z. Liu, and F. Liu, Phys. Rev. Lett. \textbf{109}, 226802 (2012).

\bibitem{Kohn1959} W. Kohn, Phys. Rev. Lett. \textbf{2}, 393-394 (1959).

\bibitem{Lazzeri2006} M. Lazzeri and F. Mauri, Phys. Rev. Lett. \textbf{97}, 266407(2006).


\bibitem{Vesta2011} K. Momma and F. Izumi, J. Appl. Crystallogr. \textbf{44}, 1272-1276 (2011).



\bibitem{Baroni2001} S. Baroni, S. de Gironcoli, and A. Dal Corso, Rev. Mod. Phys. \textbf{73}, 515-562 (2001).

\bibitem{pwscf} P. Giannozzi, \emph{et al.}, J. Phys. Condens. Matter \textbf{21}, 395502 (2009).
%


\bibitem{Troullier1991}  N. Troullier and J. L. Martins, Phys. Rev. B \textbf{43}, 1993-2006 (1991).


\bibitem{Bechstedt2012} F. Bechstedt, L. Matthes, P. Gori, and O. Pulci, App. Phys. Lett. \textbf{100}, 261906-261908 (2012).


\bibitem{me2009} J. A. Yan, W. Y. Ruan, and M. Y. Chou, Phys. Rev. B \textbf{79}, 115443 (2009).


\bibitem{Park2008} C.-H. Park, F. Giustino, M. L. Cohen, and S. G. Louie, Nano Lett. \textbf{8}, 4229-4233 (2008).

\bibitem{me2012} J. A. Yan, K. Varga, and M. Y. Chou, Phys. Rev. B \textbf{86}, 035409 (2012).

\bibitem{Ando2007} T. Ando, J. Phys. Soc. Jpn. \textbf{76}, 104711 (2007).

\bibitem{Parker1967} J. H. Parker, Jr., D. W. Feldman, and M. Ashkin, Phys. Rev. \textbf{155}, 712-714 (1967).

\bibitem{Debernardi1995} A. Debernardi, S. Baroni, and E. Molinari, Phys. Rev. Lett. \textbf{75}, 1819-1822 (1995).

\end{thebibliography}
\end{document}